\documentstyle[12pt]{article}
\begin{document}
\vskip 0.1in
\centerline{\Large\bf Negative Energies on the Brane}
\vskip .7in
\centerline{Dan N. Vollick}
\centerline{Department of Mathematics and Statistics}
\centerline{and}
\centerline{Department of Physics}
\centerline{Okanagan University College}
\centerline{3333 College Way}
\centerline{Kelowna, B.C.}
\centerline{V1V 1V7}
\vskip .9in
\centerline{\bf\large Abstract}
\vskip 0.5in
It has recently been proposed that our universe is a three-brane embedded
in a higher dimensional spacetime. Here I show that black holes on the brane,
black strings intersecting the brane, and gravitational waves propagating 
in the bulk induce an effective energy-momentum tensor on 
the brane that contains negative energy densities. 
\newpage
\section*{Introduction}
It has recently been suggested that some of the extra dimensions required by
string theory may be ``large'' \cite{Ar1,An1} or even infinite \cite{Ra1}.
In the scenario proposed in \cite{Ar1,An1} the spacetime is $M^{(4)}\times K$,
where $M^{(4)}$ is four dimensional Minkowski space and $K$ is a compact
manifold. The size of the extra dimensions must be $\stackrel{<}{\sim}$
$5\times 10^{-5}$ mm to be consistent with observations \cite{Cu1,Ha1}.
In the Randall and Sundrum model \cite{Ra1} our
three-brane is a domain wall separating two semi-infinite
anti-de Sitter regions. In both scenarios the standard model fields are
confined to the brane and gravity propagates in the bulk.
  
The Einstein field equations on the brane were derived by Shiromizu, Maeda,
and Sasaki \cite{Sh1}. The effective four dimensional
energy-momentum tensor contains
terms involving surface stresses on the brane and a term involving the
five dimensional Weyl tensor evaluated at the brane. In the absence 
of surface stresses the energy-momentum tensor on the brane reduces to 
the Weyl term which does not necessarily satisfy the weak, dominant, or 
strong energy conditions. Thus, in these ``brane-world'' scenarios the
effective four dimensional energy-momentum tensor may contain negative
energy densities.
  
The existence of negative energy densities in quantum field theory
has been known for a long time \cite{Ep1}. However, it has been shown
by Ford and Roman  (see \cite{Fo1,Fo2} for a small subset 
of their papers) that negative energy densities in Minkowski space must
satisfy quantum inequalities. For electromagnetic and scalar fields
in a four dimensional spacetime these inequalities state that an
observer can measure a negative energy density for a maximum time
$t\sim |\rho|^{-1/4}$. Thus the more negative the energy density the
shorter the time that it can persist. Quantum inequalities therefore put
constraints on the existence of negative energies in quantum field
theories in flat spacetime. 
These constraints are 
important otherwise violations of causality, cosmic censorship, and
the second law of thermodynamics could be produced.

In this paper I will show that negative energy densities are easily
produced on a three-brane embedded in a higher dimensional space.
For simplicity I will take the surface stresses on the brane to vanish
and the negative energies will originate in the Weyl part of the
effective four dimensional energy-momentum tensor. These negative
energies do not in general satisfy a quantum inequality and could
therefore present observational problems for the ``brane world'' scenarios.
   
\section*{Negative Energy Densities on a Three-Brane}
First consider a black hole of radius $r_0$ on a brane embedded in a
d-dimensional spacetime ($d>4$). Let the size of the extra dimensions
be $L$ and let $L>>r_0$. The d-dimensional metric near the black hole
is
\begin{equation}
ds^2=-f(r)dt^2+\frac{dr^2}{f(r)}+r^2d\Omega_{(d-2)}^2
\end{equation}
where
\begin{equation}
f(r)=1-\left(\frac{r_0}{r}\right)^{(d-3)}
\end{equation}
and the induced metric on the brane is
\begin{equation}
ds^2_{(4)}=-f(r)dt^2+\frac{dr^2}{f(r)}+r^2d\Omega_{(2)}^2.
\end{equation}
This is not a four dimensional Schwarzschild spacetime but it can be
interpreted as a spacetime containing a black hole and additional matter.
  
The effective four dimensional energy-momentum tenor is
\begin{equation}
T^{(4)}_{\mu\nu}=-\frac{1}{8\pi}G^{(4)}_{\mu\nu}
\label{Ein}
\end{equation}
and its non-zero components are
\begin{equation}
T^{(4)}_{tt}=-\frac{(d-4)}{8\pi r^2}\left(\frac{r_0}{r}\right)^{d-3}
\left[1-\left(\frac{r_0}{r}\right)^{d-3}\right] ,
\label{5}
\end{equation}
\begin{equation}
T^{(4)}_{rr}=\frac{(d-4)}{8\pi r^2}\left(\frac{r_0}{r}\right)^{d-3}
\left[1-\left(\frac{r_0}{r}\right)
^{d-3}\right]^{-1} ,
\end{equation}
\begin{equation}
T^{(4)}_{\theta\theta}=-\frac{(d-4)(d-3)}{16\pi}\left(\frac{r_0}{r}\right)
^{d-3} ,
\end{equation}
and $T^{(4)}_{\phi\phi}=T^{(4)}_{\theta\theta}\sin^2(\theta)$. The
energy density $\rho=-T^t_{\;\; t}$ is
\begin{equation}
\rho=-\frac{(d-4)}{8\pi r^2}\left(\frac{r_0}{r}\right)^{d-3} 
\label{8}
\end{equation}
and the brane therefore contains a negative energy density.  It is 
important to remember that (\ref{5}) to (\ref{8}) are valid only for 
$r<<L$. For $r>>L$ the spacetime becomes approximately Schwarzschild 
and $f(r)\simeq 1-2m/r$ where $m$ is the four dimensional
asymptotic mass. It was shown in \cite{Em1} that this mass is the 
same as the mass measured in the bulk. The negative mass contained
within a radius $0<R<<L$ can be found from \cite{Mi1}
\begin{equation}
m(R)=4\pi\int_0^Rr^2\rho(r)dr
\end{equation}
which diverges. Thus there is an infinite amount of negative mass
in the spacetime.
   
In five dimensions the above energy-momentum tensor can also be found
using the approach of Shiromizu et. al. \cite{Sh1}. Using the 
Gauss-Codacci
equations they show that in the absence of surface stresses
and a cosmological constant the Einstein tensor is given by
\begin{equation}
G^{(4)}_{\mu\nu}=-C^{(5)}_{\alpha\beta\rho\sigma}\eta^{\alpha}\eta^{\rho}
q_{\mu}^{\;\;\beta}q_{\nu}^{\;\;\sigma}
\label{Weyl}
\end{equation}
where $C^{(5)}_{\alpha\beta\rho\sigma}$ is the five dimensional
Weyl tensor evaluated at the brane, $\eta^{\alpha}$ is a unit vector
orthogonal to the brane, and $q_{\mu\nu}=g_{\mu\nu}-\eta_{\mu}\eta_{\nu}$
is the induced metric on the brane. The five dimensional metric is
\begin{equation}
ds^2=-f(r)dt^2+\frac{dr^2}{f(r)}+r^2[d\chi^2+\sin^2\chi d\Omega^2_{(2)}]
\end{equation}
and the brane is located at $\chi=\pi/2$. The energy-momentum tensor
on the brane is
\begin{equation}
T^{(4)}_{\mu\nu}=\frac{1}{8\pi r^2}C^{(5)}_{5\mu 5\nu} 
\label{EM}
\end{equation}
and the relevant non-zero components of the Weyl tensor at $\chi=\pi/2$ are
\begin{equation}
C^{(5)}_{5151}=-\left(\frac{r_0}{r}\right)^2\left[1-\left(\frac{r_0}{r}\right)^2
\right] ,
\end{equation}
\begin{equation}
C^{(5)}_{5252}=\left(\frac{r_0}{r}\right)^2
\left[1-\left(\frac{r_0}{r}\right)^2\right]^{-1},
\end{equation}
\begin{equation}
C^{(5)}_{5353}=-r_0^2 ,
\end{equation}
and
\begin{equation}
C^{(5)}_{5454}=-r_0^2\sin^2(\theta) .
\end{equation}
Substituting these into (\ref{Ein}) and (\ref{Weyl})
gives the same energy-momentum tensor
as found above.

A neutral black string wrapped around the extra dimension can also
produce negative energy densities on the brane. Such a black string in
a d-dimensional spacetime can be obtained by taking the product of 
$S^1$ with a $(d-1)$-dimensional Schwarzschild spacetime. The induced
metric on the brane near the string will be
\begin{equation}
ds_{(4)}^2=-g(r)dt^2+\frac{dr^2}{g(r)}+r^2d\Omega_{(2)}^2
\end{equation}
where
\begin{equation}
g(r)=1-\left(\frac{r_0}{r}\right)^{d-4} .
\end{equation}
and $r_0<<L$.
Thus, if $d>5$ the brane will contain negative energy densities.
  
Negative energies can also be produced on the brane by a gravitational
wave propagating in the bulk. Consider the five dimensional spacetime
\begin{equation}
ds^2=-dt^2+[1+\epsilon\cos k(t-z)]dx^2+dy^2+dz^2+2\epsilon
\cos k(t-z)dxdw+[1-\epsilon\cos k(t-z)]dw^2
\end{equation}
that contains a linearized gravitational wave propagating in the z-direction.
The fifth dimension is topologically $S^1$ and is labelled by $w$.
The brane is located at $w=0$ and the effective four dimensional 
energy-momentum tensor, to lowest order in $\epsilon$, is
\begin{equation}
T^{(4)}_{tt}=\frac{\epsilon k^2}{16\pi}\cos k(t-z) ,
\end{equation}
\begin{equation}
T^{(4)}_{xx}=T^{(4)}_{yy}=0 ,
\end{equation}
\begin{equation}
T^{(4)}_{zz}=\frac{\epsilon k^2}{16\pi}\cos k(t-z) ,
\end{equation}
and
\begin{equation}
T^{(4)}_{tz}=T^{(4)}_{zt}=-\frac{\epsilon k^2}{16\pi}\cos k(t-z) .
\end{equation}
The energy density on the brane, to lowest order in $\epsilon$, is
\begin{equation}
\rho=\frac{\epsilon k^2}{16\pi}\cos k(t-z)
\end{equation}
and consists of regions of positive and negative energy densities propagating
at the speed of light. Note that $\rho$ is proportional to $\epsilon$ in
contrast to the (positive) energy density 
in a four dimensional gravitational wave, 
which is proportional to $\epsilon^2$.
The same energy-momentum tensor can be obtained
using (\ref{Ein}) and (\ref{Weyl}).
    
Finally, consider a gravitational wave incident upon the brane. The metric
will be taken to be
\begin{equation}
ds^2=-dt^2+[1+\epsilon\cos k(t-w)]dx^2+[1-\epsilon\cos k(t-w)]dy^2+
2\epsilon\cos k(t-w)dxdy +dz^2+dy^2 .
\end{equation}
For the spacetime to be periodic in the fifth dimension $k$ must
satisfy $k=2n\pi/L$, where $n$ is an integer.
The effective four dimensional energy-momentum tensor, to lowest order in
$\epsilon$, is
\begin{equation}
T^{(4)}_{tt}=T^{(4)}_{zz}=0 ,
\end{equation}
\begin{equation}
T^{(4)}_{xx}=-\frac{\epsilon k^2}{16\pi}\cos(kt)  ,
\end{equation}
\begin{equation}
T^{(4)}_{yy}=\frac{\epsilon k^2}{16\pi}\cos(kt) ,
\end{equation}
\begin{equation}
T^{(4)}_{xy}=T^{(4)}_{yx}=-\frac{\epsilon k^2}{16\pi}\cos(kt) .
\end{equation}
This energy-momentum tensor violates the weak, strong, and dominant energy
conditions. Thus, even though $\rho=0$ in these coordinates some observers
on the brane will measure negative energy densities.
\section*{Conclusion}
I have shown that black holes on the brane, black strings intersecting
the brane, and gravitational waves propagating in the bulk produce
negative energy densities on the brane. 
The total negative energy around the black hole and black string
diverges ($d>5$ for the string) and the negative energy density produced by 
a gravitational wave propagating in the bulk is 
proportional to the wave amplitude,
not the amplitude squared.


\begin{thebibliography}{1}
\bibitem{Ar1}
N. Arkani-Hamed, S. Dimopoulos, and G. Dvali, Phys. Lett. {\bf B429},
263 (1998), hep-ph/9803315
\bibitem{An1}
I. Antoniadis, N. Arkani-Hamed, S. Dimopoulos, and G. Dvali, Phys. Lett.
{\bf B436}, 257 (1998), hep-ph/9804398
\bibitem{Ra1}
L. Randall and R. Sundrum, hep-th/9906064
\bibitem{Cu1}
S. Cullen and M. Perelstein, Phys. Rev. Lett. {\bf 83}, 268 (1999),
hep-ph/9903422
\bibitem{Ha1}
L.J. Hall and D. Smith, Phys. Rev. {\bf D60}, 085008 (1999), 
hep-ph/9904267
\bibitem{Sh1}
T. Shiromizu, K. Maeda, and M. Sasaki, gr-qc/9910076
\bibitem{Ep1}
H. Epstein, V. Glaser, and A. Jaffe, Nouvo Cimento {\bf 36}, 1016 (1965)
\bibitem{Fo1}
L.H. Ford and T.A. Roman, Phys. Rev {\bf D55}, 2082 (1997)
\bibitem{Fo2}
L.H. Ford, Phys. Rev. {\bf D43}, 3972 (1991)
\bibitem{Em1}
R. Emparan, G.T. Horowitz, R.C. Myers, hep-th/0003118
\bibitem{Mi1}
C.W. Misner, K.S. Thorne, and J.A. Wheeler, Gravitation (San Francisco:
Freeman) 1973
\end{thebibliography}
\end{document}